\documentclass[nofootinbib,preprintnumbers,floats,twocolumn,prd,aps,superscriptaddress]{revtex4}
\usepackage{latexsym}
\usepackage{amsfonts}
\usepackage{psfrag}
\usepackage{graphicx}
\usepackage{bm}
\usepackage{amssymb,amsmath,epsfig}
\usepackage{subfig}
\usepackage{float}
\usepackage[titletoc,title]{appendix}

\begin{document}

\setlength{\pdfpagewidth}{8.5in}
\setlength{\pdfpageheight}{11in}
\renewcommand{\thefootnote}{\arabic{footnote}}

\title{Inflying perspectives of reduced phase space}
%Wavepacket Scattering with Non-inertial Observer Networks}
%Scattering Wavepackets with Non-inertial Reference (or Coordinate) Frames
%\title{Frame Effects in Reduced Phase Space Scattering}
%Observer Dependence in Reduced Phase Space Quantization
%Scattering with Non-inertial Observer Networks

\author{Cisco Gooding}%\email{cgooding@physics.ubc.ca}
\affiliation{
	School of Mathematical Sciences, 
	University of Nottingham, UK
	}
\affiliation{
  Department of Physics \& Astronomy,
  University of British Columbia, Canada
}
\author{William G. Unruh}
\affiliation{
  Department of Physics \& Astronomy,
  University of British Columbia, Canada
}
\date{\today}

\begin{abstract}
There is widespread disagreement about how the general covariance of a theory affects its quantization. Without a complete quantum theory of gravity, one can examine quantum consequences of coordinate choices only in highly idealized `toy' models. In this work, we extend our previous analysis of a self-gravitating shell model \cite{GoodingUnruh14}, and demonstrate that coordinate freedom can be retained in a reduced phase space description of the system. We first consider a family of coordinate systems discussed by Martel and Poisson \cite{PainleveFamily}, which have time coordinates that coincide with the proper times of ingoing and outgoing geodesics (for concreteness, we only consider the former). Included in this family are Painlev\'e-Gullstrand coordinates, related to a network of infalling observers that are asymptotically at rest, and Eddington-Finkelstein coordinates, related to a network of infalling observers that travel at the speed of light. We then introduce \textit{inflying} coordinates - a hybrid coordinate system that allows the infalling observers to be arbitrarily boosted from one member of the aforementioned family to another. We perform a phase space reduction using inflying coordinates with an unspecified boosting function, resulting in a reduced theory with residual coordinate freedom. Finally, we discuss quantization, and comment on the utility of the reduced system for the study of coordinate effects and the role of observers in quantum gravity. 
\end{abstract}

\maketitle

\section{Introduction}

The question of how to reconcile general covariance with Dirac's definition of observables is profoundly unsettled, and poses one of the most perplexing obstacles to unifying gravity and the quantum. If observables are to be invariant under diffeomorphism, then the local observables of quantum field theory are no longer viable, and one is hard-pressed to identify more than the most trivial global charges as an observable. On the other hand, how could observables break diffeomorphism symmetry while still maintaining consistency with general relativity?

In previous work \cite{GoodingUnruh14}, we studied the quantization of a reduced phase space description of a spherical, infinitesimally thin, self-gravitating shell. The phase space reduction employed a specific coordinate system; namely, the well-known stationary, nonstatic coordinate system for the Schwarzschild geometry, known as ``Painlev\'e-Gullstrand'' coordinates. This choice was made largely to simplify the analysis, but also for the attractive features of the coordinates themselves, such as flat spatial slices, horizon penetration, and compactness of form. One naturally wonders, however, which aspects of the reduced phase space description (and its quantization) are influenced by this choice. The canonical momentum in the reduced system certainly depends on the coordinate choice, but one can show that a broad set of choices lead to the same reduced classical action \cite{Menotti}.

Though our ultimate goal is to investigate how coordinates used to carry out a phase space reduction influence quantization, the material presented here is primarily classical. We present a reduced model for the self-gravitating shell that has residual coordinate freedom, and describe how the reduced model can be used to compare representations of quantum wavepacket scattering based on different coordinate choices. 

The coordinates we use are constructed from a family of coordinate systems, with each member associated with a network of infalling geodesic observers; given an initial velocity at spatial infinity, and corresponding infalling geodesic, there is a unique member of the coordinate family with a time coordinate that coincides with the proper time of the infalling geodesic. The Painlev\'e-Gullstrand coordinates are therefore coordinate family members, associated with infalling observers that have vanishing asymptotic velocity. We then subject this observer network to local boosts, and examine the effect boosting has on the reduced phase space of our model system. 

This paper is organized as follows. In Section \ref{CoordIntro}, we review the set of coordinate systems studied by Martel and Poisson \cite{PainleveFamily}, which generalize Painlev\'e-Gullstrand coordinates. This set will serve as our coordinate family (hereafter referred to as the Painlev\'e-Gullstrand family). In Section \ref{Sec:Hybrid}, we introduce ``inflying'' coordinates, a hybridization of the Painlev\'e-Gullstrand family, constructed by the authors to allow arbitrary local boosting to the associated observer network.

The canonical structure of the classical self-gravitating shell model is defined at the kinematic level in Section \ref{Sec:Action}, and in Section \ref{Sec:PhaseReductionPainleve} we use the inflying coordinates to derive a reduced phase space representation. Asymptotics of the reduced Hamiltonian are given in Section \ref{Sec:Asymptotics}. We then take preliminary steps towards quantization of the model in Section \ref{Sec:Quantization}, and are confronted by difficulties obtaining a consistent probabilistic interpretation outside the nonrelativistic regime. The source of the problem is explained in Section \ref{Sec:Discussion}. We conclude by discussing future work to resolve the probability issue encountered here, and implications of our results for developments in quantum gravity.

\section{Coordinate Families and Observer Networks}
\label{CoordIntro}

The Painlev\'e-Gullstrand coordinate family used here was previously studied by Martel and Poisson \cite{PainleveFamily}. We will express this family using the Arnowitt-Deser-Misner (ADM) form of the metric \cite{ADM}. In spherical symmetry, the ADM metric can be written
\begin{equation}
g_{\mu\nu}dx^{\mu}dx^{\nu}=-N^2 dt^2 + L^2\left(dr+N^r dt\right)^2+R^2d\Omega^2,
\label{eq:Metric}
\end{equation}
where $N$ is the lapse function, $N^r$ is the radial component of the shift vector, and $L^2$ and $R^2$ are the only independent components of the spatial metric. In terms of these metric variables, the Painlev\'e-Gullstrand family members are written
\begin{equation}
\hspace{8pt} L = \lambda, \hspace{8pt} R=r, \hspace{8pt} N = \frac{1}{\lambda}, \hspace{8pt} N^r=\pm\frac{\sqrt{1-\lambda^2 f}}{\lambda^2},
\label{eq:PainleveFam}
\end{equation}
with $f=1-\frac{2\mathcal{M}}{r}$, $\mathcal{M}$ being the (enclosed) ADM mass of the spacetime, and $0<\lambda\leq 1$ parametrizing the family. In the $\lambda\rightarrow 0$ limit, the coordinates defined by (\ref{eq:PainleveFam}) become the familiar (null) Eddington-Finkelstein coordinates (also known as Penrose-Eddington-Finkelstein coordinates, due to Penrose's explicit initial use of them \cite{Penrose65}). In the upper limit, $\lambda\rightarrow 1$, the coordinates (\ref{eq:PainleveFam}) reduce to the Painlev\'e-Gullstrand coordinates used in our previous work \cite{GoodingUnruh14}. 

Taking the positive sign for the shift vector $N^r$, the line element takes the form
\begin{equation}
ds^2=\frac{1}{\lambda^2}dt_\lambda^2+\lambda^2\left(dr+\frac{1}{\lambda^2}\sqrt{1-\lambda^2 f}dt_\lambda\right)^2+r^2d\Omega^2.
\label{line}
\end{equation}
This family of coordinate systems still possesses the connection between coordinate lines and infalling observers, as with the ($\lambda=1$) Painlev\'e-Gullstrand coordinates: for a geodesic observer falling radially inward, the proper time along the trajectory is equal to the Painlev\'e-Gullstrand time if the observer starts from rest at infinity, and equal to the time coordinate of another member of the PG family if the observer starts with a nonzero velocity. The quantity $\lambda$ is related to the ``initial velocity at infinity''\footnotemark \hspace{4pt}of the infalling observers:
\begin{equation}
\lambda = \sqrt{1-v_\infty^2}.
\end{equation}
\footnotetext{Following the convention from \cite{PainleveFamily}, we define positive observer velocity to be radially inward, such that $v_\infty$ takes on values from $0$ to $1$ as $\lambda$ varies from $1$ to $0$. This is the opposite convention as the one used for the shell velocity.} The initial velocity is in turn related to the geodesic observer's energy per unit rest mass (also known as the relativistic ``gamma'' factor), through the standard expression $\gamma=1/\sqrt{1-v_\infty^2}$ \cite{PainleveFamily}. 

A particular member of the Painlev\'e-Gullstrand family can therefore be associated with a network of geodesic observers, each with the same energy per unit rest mass. Any such network has the property that its observers all have proper $4$-velocities equal to a constant times the gradient of a time function $t_\lambda$, with the time function given by
\begin{equation}
t_\lambda=T+\int{dr\, \frac{\sqrt{1-\lambda^2 f}}{f}}.
\label{eq:TimeTransFS}
\end{equation}
Here $T$ is the Schwarzschild time, which is related to the Painlev\'e-Gullstrand time $t$ by
\begin{equation}\label{PainleveTime}
t=T+4\mathcal{M}\left(\sqrt{\frac{r}{2\mathcal{M}}}+\frac{1}{2}\ln{\left|\frac{\sqrt{\frac{r}{2\mathcal{M}}}-1}{\sqrt{\frac{r}{2\mathcal{M}}}+1}\right|}\right).
\end{equation}

\section{Inflying Coordinates}
\label{Sec:Hybrid}

To compare quantizations that result from different coordinate choices, we would like to combine members of the Painlev\'e-Gullstrand family into a hybrid coordinate system, such that the associated observer network is made of infalling observers with different energy to mass ratios. We do this for the following reason: to define an initial quantum state in the same way for different coordinate choices, it is useful to have a region of spacetime where the different coordinates are equivalent. This way, for each of the spatial hypersurfaces defined by the different coordinate choices, there is a region that is locally equivalent with which to define an initial quantum state. Otherwise, a definition of initial time for one coordinate choice would correspond to a range of times for a different coordinate choice, making the preparation of identical initial states a difficult endeavour indeed.

The plan is then to build a hybrid coordinate system, and use it to compare the evolution of an initial state with the evolution of the same initial state determined by the quantization based on a member of the Painlev\'e-Gullstrand family of coordinates that shares a portion of its initial time hypersurface with the hybrid coordinates.

Up to this point, the networks we have described are composed entirely of inertial observers. Once we hybridize the Painlev\'e-Gullstrand family members, the associated observer networks will be locally boosted, and non-inertial features will arise. To construct the hybrid coordinates, we will use an ansatz for the spatial metric components that resembles the coordinate choice of an arbitrary Painlev\'e-Gullstrand family member:
\begin{equation}\label{HybridAnsatz}
\hspace{8pt} L = \lambda\left(r\right), \hspace{18pt} R=r
\end{equation}
The only difference is that now we allow $\lambda$ to be an as-yet unspecified function of the areal radius, $r$. If we insert the ansatz (\ref{HybridAnsatz}) into the gravitational equations of motion, and make use of the Hamiltonian and momentum constraint, we find
\begin{equation}
N = \frac{1}{\lambda(r)}, \hspace{18pt} N^r=\pm\frac{\sqrt{1-\lambda(r)^2 f}}{\lambda(r)^2}.
\label{eq:PainleveHybrid}
\end{equation}
Remarkably, the hybrid metric we arrive at has the exact form of the Painlev\'e-Gullstrand family of coordinates, except instead of having the freedom to specify a continuous parameter $\lambda$, we now have the freedom to choose any function $\lambda(r)$ satisfying $0<\lambda(r)\leq 1$. As $\lambda$ varies with $r$, the associated observers get boosted arbitrarily, producing deviations from the inertial ``infalling'' that occurs for each member of the Painlev\'e-Gullstrand family; we will therefore refer to these hybrid coordinates as \textit{inflying}.

\section{Action for a Self-gravitating Shell Model}
\label{Sec:Action}

So far we have only considered pure spherically-symmetric gravity, in the absence of matter. We will add matter to our universe in a simple way, using a model of a self-gravitating fluid shell introduced by the authors \cite{GoodingUnruh14}. The fluid shell is infinitesimally thin, and has tangential pressure on its surface, with an equation of state parametrized by a mass function $M(\hat{R})=\hat{M}$ that depends on the areal radius of the shell (a hat denotes that a quantity is to be evaluated on the shell). 

In Hamiltonian form, the action for this model, including gravity, is given by
\begin{equation}\label{Action}
I = \int{dt\, p\dot{x}}+\int{dt\,dr\,\left(\pi_R \dot{R}+\pi_L \dot{L}-NH_0-N^r H_r\right)},
\end{equation}
such that
\begin{eqnarray}
H_0 &=& \frac{L\pi_L^2}{2R^2}-\frac{\pi_L\pi_R}{R}+\left(\frac{R R'}{L}\right)'-\frac{(R')^2}{2L}-\frac{L}{2} \nonumber\\
&& +\sqrt{L^{-2}p^2+M(R)^2}\delta(r-x),\nonumber\\
H_r &=& R'\pi_R-L\pi_L'-p\delta(r-x).
\label{eq:GravConstraints}
\end{eqnarray}
In these expressions, a prime indicates differentiation with respect to $r$, and an overdot denotes differentiation with respect to $t$. $x(t)$ is the areal radius of the shell at time $t$,  and $p$, $\pi_R$, and $\pi_L$ are the momenta conjugate to $x$, $R$, and $L$ (respectively) in the kinematical phase space. Variations with respect $x$, $R$, $L$, and their conjugate momenta yield the equations of motion for the system. Variations with respect to the lapse $N$ and shift $N^r$ produce $H_0=0$ and $H_r=0$ (respectively), which are known as the Hamiltonian and momentum constraints.

\section{Phase Space Reduction in Inflying Coordinates}
\label{Sec:PhaseReductionPainleve}

The full phase space $\Gamma$ of the system defined by (\ref{Action}) is called the kinematical phase space. Along with the shell variable $x$, $\Gamma$ has two gravitational field variables ($L$ and $R$), and produces two gravitational constraints ($H_0=0$ and $H_r=0$). Consequently, if one solves the constraints for the gravitational momenta and inserts those solutions back into the original action, the resulting system will no longer have gravitational freedom. This approach is called phase space reduction, and when applied to our shell-plus-gravity system leads to a reduced phase space that depends only on the shell variable $x$ (and its conjugate momentum). 

The reduced phase space $\bar{\Gamma}$ is defined as the set of equivalence classes in $\Gamma$ under changes of coordinates. Working with equivalence classes can be cumbersome, depending on what calculations one wants to perform. Fortunately, it suffices to select a representative from each equivalence class by choosing a coordinate system, provided one respects the canonical structure of $\Gamma$. We now present the phase space reduction associated with our inflying coordinates.

The starting point for the phase space reduction parallels the reduction presented in \cite{GoodingUnruh14}, which was based on the Painlev\'e-Gullstrand coordinate choice. We first consider the Liouville form $\mathcal{F}$ on the full (kinematical) phase space $\Gamma$, which is given by
\begin{eqnarray}
\mathcal{F}=p \bm{\delta} x + \int{dr\left(\pi_L \bm{\delta} L +\pi_R \bm{\delta} R\right)}.
\end{eqnarray}
We can then pull this back to the representative hypersurface $\bar{H}_\lambda \subseteq \Gamma$ associated with the particular coordinate choice for a given $\lambda(r)$. The pulled-back Liouville form $\mathcal{F}_\lambda$ induces a Liouville form on the reduced phase space through the isomorphism between the reduced phase space $\bar{\Gamma}$ and the representative hypersurface $\bar{H}_\lambda \subseteq \Gamma$ (for more details, see \cite{GoodingUnruh14,Louko}). The induced Liouville form on $\bar{\Gamma}$ yields the canonical structure of our reduced system. 

Away from the shell, take the following linear combination of the constraints:
\begin{equation}
-\frac{R'}{L}H_0-\frac{\pi_L}{RL}H_r=\mathcal{M}',
\end{equation}
for
\begin{equation}\label{ADMMassFunction}
\mathcal{M}(r)=\frac{\pi_L^2}{2R}+\frac{R}{2}-\frac{R(R')^2}{2L^2}.
\end{equation}
The quantity $\mathcal{M}(r)$ corresponds to the ADM mass $H$ when evaluated outside of the shell, and vanishes inside the shell. Away from the shell, (\ref{ADMMassFunction}) enables us to solve for $\pi_L$, and the momentum constraint $H_r=0$ then yields $\pi_R$. In our new coordinates (\ref{eq:PainleveFam}), these gravitational momenta solutions are given by
\begin{equation}
\pi_L=\pm R\sqrt{\left(\frac{R'}{\lambda}\right)^2-1+\frac{2\mathcal{M}(r)}{R}},\hspace{8pt}\pi_R=\frac{\lambda}{R}\pi_L'.
\label{eq:GravMomentaF}
\end{equation}

As with our previous (Painlev\'e-Gullstrand) coordinate choice \cite{GoodingUnruh14}, we must take care to include a deformation region in $R$, near the shell ($x-\epsilon<r<x$), in order to satisfy the gravitational constraints.

We can determine what conditions the gravitational constraints impose on the metric function $R$ by integrating these constraints across the shell, and assuming both continuity of the (spatial) metric and finiteness of the gravitational momenta. One then finds the conditions
\begin{equation}
\Delta R'=-\frac{\bar{V}}{\hat{R}}, \hspace{8pt} \Delta\pi_L=-\frac{p}{\lambda},
\label{eq:JumpsF}
\end{equation}
where $\bar{V}=\sqrt{p^2+\bar{M}^2}$, $\bar{M}=\hat{M}\lambda$, and $\Delta$ indicates the jump of a quantity across the shell. By inspection, the metric function $R$ defined in \cite{GoodingUnruh14} can be generalized as
\begin{equation}
R(r,t)=r-\frac{\epsilon}{x}\bar{V}g\left(\frac{x-r}{\epsilon}\right),
\label{eq:CoordsF}
\end{equation}
for a function $g$ having the properties 
\begin{align}
 \lim_{z\rightarrow 0^+}g'(z)&=1\\
 \lim_{z\rightarrow 0^-}g'(z)&=0\,,
\end{align}
from which follows
\begin{align}
&\lim_{\epsilon\rightarrow 0}R' (x-\epsilon)=1+\frac{\bar{V}}{x}\\
&\lim_{\epsilon\rightarrow 0}R' (x+\epsilon)=1\,.
\end{align}

As can be expected from the generally nondynamical form of our coordinate choice, the $\pi_R$ term integrated over the deformation region will give the only contribution to the pullback of the Liouville form:
\begin{equation}
\mathcal{F}_\lambda=p\bm{\delta} x+
\int_{x-\epsilon}^x dr\, \pi_R \bm{\delta} R.
\end{equation} 
In the $\epsilon\rightarrow0$ limit, we have, in the deformation region,
\begin{equation}
\pi_R = \frac{x R''}{\lambda\sqrt{\left(\frac{R'}{\lambda} \right)^2-1}}+\mathcal{O}(1),
\end{equation}
which allows us to express the gravitational contribution to the Liouville form as
\begin{equation}
\int_{x-\epsilon}^{x}{dr\, \pi_R \bm{\delta} R}=x\bm{\delta} x\int_{x-\epsilon}^{x}{dr\,\frac{R'' \left(1-R'\right)}{\lambda\sqrt{\left(\frac{R'}{\lambda}\right)^2-1}}}+\mathcal{O}(\epsilon).
\end{equation}
Here we are assuming that variations of $\lambda$ in the deformation region can be neglected in the $\epsilon\rightarrow0$ limit. We can then change the integration variable from $r$ to $v=R'$, which yields
\begin{equation}
\int_{x-\epsilon}^{x}{dr\, \pi_R \bm{\delta} R}=\frac{x\bm{\delta} x}{\lambda}\int_{1}^{R_-'}{dv\,\frac{\left(1-v\right)}{\sqrt{\left(\frac{v}{\lambda}\right)^2-1}}}+\mathcal{O}(\epsilon),
\end{equation}
with $R_-'$ being $R'$ evaluated just inside the shell. Integrating and rearranging then leads to
\begin{eqnarray}
\bm{\delta} x \bigg[&-P-x\lambda\sqrt{w}+x\left(\sqrt{1-\lambda^2}-\ln{\left(1+\sqrt{1-\lambda^2}\right)}\right)\nonumber\\
&+x\ln{\left(1+\lambda\sqrt{w}+\frac{\bar{V}+P}{x}\right)}\bigg]
\end{eqnarray}
(plus terms that vanish as $\epsilon\rightarrow 0$), where we have used the definition
\begin{equation}
w \equiv \frac{1}{\lambda^2}-1+\frac{2H}{x}.
\end{equation}
This completes the calculation of $\mathcal{F}_\lambda$, the pullback of the full Liouville form $\mathcal{F}$ to $\bar{H}_\lambda$:
\begin{equation}
\mathcal{F}_\lambda=p_\lambda \bm{\delta} x,
\end{equation}
with the reduced canonical momentum evidently given by
\begin{eqnarray}
p_\lambda = &-&x\lambda\sqrt{w}+ x\left[ \sqrt{1-\lambda^2}-\ln{\left(1+\sqrt{1-\lambda^2}\right)}\right]  \nonumber\\
&+&x\ln{\left(1+\frac{\bar{V}+p}{x}+\lambda\sqrt{w}\right)}.
\label{eq:PcF}
\end{eqnarray}
In the reduced phase space, the unreduced momentum $p$ becomes a constrained function of $H$ and $x$. One can obtain this function by inserting the gravitational momentum solutions away from the shell given by equation (\ref{eq:GravMomentaF}) into the jump equations (\ref{eq:JumpsF}) and squaring. We then find that $p$ is constrained to obey
\begin{equation}
\lambda^2 H = \sqrt{p^2+\hat{M}^2\lambda^2}+\frac{\bar{M}^2}{2 x}-p\lambda\sqrt{w}.
\label{eq:DeterminesPF}
\end{equation}
Fortunately, equation (\ref{eq:DeterminesPF}) can easily be transformed into a quadratic, which has the explicit solution 
\begin{eqnarray}
p = \frac{\sqrt{1-\lambda^2 f}h\pm\sqrt{h^2-M^2 f}}{f},
\label{eq:PexplicitF}
\end{eqnarray}
with $f=1-2H/x$ and $h=H-M^2/2x$. Note that although we use a $\lambda$ subscript to distinguish the reduced momentum (\ref{eq:PcF}) from the kinematical momentum (\ref{eq:PexplicitF}), both quantities depend on the choice of $\lambda$.

Due to the second derivatives present in the Einstein-Hilbert action, a nonzero boundary variation results from integrating by parts the term $\int{dt_\lambda\, dr\,N^rL(\delta \pi_L)'}$, which is part of the momentum constraint. With standard Painlev\'e-Gullstrand coordinates, the boundary term is (very conveniently) equal to $-\int{dt\, H}$, with $H$ being the ADM mass. One can then identify the ADM mass with the reduced Hamiltonian of the system, generating asymptotic time translations. We will now demonstrate that this property holds for an arbitrary inflying coordinate choice.

The boundary in consideration is spatial infinity, and as $r\rightarrow \infty$ we now have $N^r \rightarrow \sqrt{1-\lambda^2 f}/\lambda^2$, $N \rightarrow 1/\lambda$, and $\pi_L\rightarrow r \sqrt{1/\lambda^2-1+2\mathcal{M}/r}$. It then readily follows that 
\begin{equation}
\delta (\pi_L)\rightarrow \frac{\delta H}{\sqrt{\frac{1}{\lambda^2}-1+\frac{2\mathcal{M}}{r}}},
\end{equation}
and so $\delta (\pi_L)N^r L\rightarrow \delta H$. From this we can conclude that the variation of the boundary term is canceled if we add to the action the term
\begin{equation}
I_{bdry}=-\int{dt_\lambda\, H},
\end{equation}
which has the same form as before, with the previous time $t$ replaced by our new time coordinate $t_\lambda$.

We can then determine the reduced action associated with each family member by adding the boundary term to the action defined by $\mathcal{F}_\lambda$. The result is
\begin{equation}
I_{\lambda}=\int{dt_\lambda\, \left(p_\lambda \frac{dx}{dt_\lambda} - H\right)},
\label{eq:ReducedActionF}
\end{equation}
with the reduced momentum now given by (\ref{eq:PcF}). We thus conclude from the form of the reduced action (\ref{eq:ReducedActionF}) that the ADM mass is the reduced Hamiltonian for the system, regardless of the coordinate hybridization.

\section{Weak-field and Flat Spacetime Limits}\label{Sec:Asymptotics}

We now consider the asymptotic structure of the reduced system defined by (\ref{eq:ReducedActionF}), in the weak-field limit. Though we work in natural units $c=G=1$, the weak-field limit corresponds to $G\rightarrow 0$; for slowly-varying boosting functions $\lambda(r)$, we can equivalently take $x\rightarrow\infty$.

In the limit of flat spacetime, (\ref{eq:PcF}) becomes
\begin{equation}\label{eq:PExact}
p_\lambda = H\sqrt{1-\hat{\lambda}^2}\pm\sqrt{H^2-\hat{M}^2}.
\end{equation} 
Since $\hat{\lambda}$ is related to the local velocity attributed to the observer network through the expression $\hat{v}=\sqrt{1-\hat{\lambda}^2}$, (\ref{eq:PExact}) is just as one might expect: the shell momentum defined with respect to our coordinate system is given by the usual relativistic expression, offset by a momentum $H\hat{v}$ that is attributed to the shell due to the infalling nature of the coordinates. In the nonrelativistic limit, the offset becomes the familiar $\hat{M}\hat{v}$.

The shell is not subject to self-gravitation in this limit, but the dynamics still includes tangential pressure (produced by nonconstant regions of $\hat{M}=M(x)$) and the influence of boosts (produced by nonconstant regions of $\hat{\lambda}=\lambda(x)$). Taking $G\rightarrow 0$ also implies $p_\lambda=p$, so we will drop the $\lambda$ subscript on the momentum, keeping in mind that the definitions of both $p$ and $p_\lambda$ depend on $\lambda$. The Hamiltonian defined by (\ref{eq:PExact}) is then given by
\begin{align}\label{Hflat}
H=\frac{1}{\hat{\lambda}^2}\left(-p\sqrt{1-\hat{\lambda}^2}+\sqrt{p^2+\hat{M}^2\hat{\lambda}^2}\right).
\end{align}

It will be helpful to make contact with familiar results from classical mechanics, so we will also apply our approach to the simpler case of a Hamiltonian that is quadratic in momentum. The general form of such a Hamiltonian is
\begin{eqnarray}\label{QuadHamilt}
H(x,p)\approx H_0(x)+H_1(x) p+H_2(x) p^2,
\end{eqnarray}
which we will use as an ansatz for our weak-field Hamiltonian. As a reference, the specific functions $\{H_i\}$ for the flat spacetime limit can be found by expanding the Hamiltonian (\ref{Hflat}) to second order in $p$, giving 
\begin{equation}
H_0(x)=\frac{\hat{M}}{\hat{\lambda}}, \hspace{3pt}
H_1(x)=-\frac{\sqrt{1-\hat{\lambda}^2}}{\hat{\lambda}^2},\hspace{3pt}
H_2(x)=\frac{1}{2\hat{M}\hat{\lambda}^3}.
\end{equation}

Though it is not possible to solve the relation 
(\ref{eq:PcF}) for the Hamiltonian $H$ when $x$ is finite, we can obtain an approximate Hamiltonian for large $x$ and small momentum $p_\lambda$. We begin by choosing the ansatz
\begin{eqnarray}\label{ArbHamF}
H=H_0+H_1 p_\lambda+H_2 p_\lambda^2,
\end{eqnarray}
with some functions $H_0(x)$, $H_1(x)$, and $H_2(x)$.
We substitute this ansatz into the relation (\ref{eq:PcF})
and expand for large $x$ and small $p_\lambda$.
Then we compare the coefficients for the various powers 
of $p_\lambda$ and deduce for $\lambda=1$ 
\begin{eqnarray}
H_0&\sim&M-\frac{M^2}{18 x}+\frac{2M^3}{405 x^2}\\
H_1&\sim&-\frac{2}{3}\sqrt{\frac{2M}{x}}-\frac{M^{3/2}}{135\sqrt{2}x^{3/2}}+\frac{161M^{5/2}}{48600\sqrt{2}x^{5/2}}\\
H_2&\sim&\frac{1}{2M}+\frac{1}{3x}+\frac{M}{270x^2}.
\end{eqnarray}
When the coordinate system corresponds to an observer network with
finite ``velocity at infinity,'' we find
\begin{eqnarray}
\label{H0F}H_0&\sim&\frac{M}{\lambda}-\frac{\lambda M^3}{24(1-\lambda^2)x^2}\\
\label{H1F}H_1&\sim&-\frac{\sqrt{1-\lambda^2}}{\lambda^2}-\frac{M}{2x\lambda\sqrt{1-\lambda^2}}+\frac{M^2}{6(1-\lambda^2)^{\frac{3}{2}}x^2}\\
\label{H2F}H_2&\sim&\frac{1}{2M\lambda^3}+\frac{1}{2x\lambda^2}-\frac{3M}{16x^2 \lambda\left(1-\lambda^2\right)}.
\end{eqnarray}
Note that these asymptotic expansions do not 
have the correct limit for $\lambda\rightarrow 1$.
The reason for this is that for large $x$ and $\lambda<1$, we have $\sqrt{1-\lambda^2 f}\sim 1/x$ (modulo a constant)
whereas for $\lambda=1$, we find $\sqrt{1- f}\sim 1/\sqrt{x}$.
We mention that it is possible to find to the order to which we expanded the Hamiltonian an interpolating function, though we will not require such a construction for the purposes of this paper.

The Hamiltonian expansions given above should only be used in regions for which $\lambda(x)$ and $M(x)$ vary slowly compared to the effective gravitational potentials. 

\section{Quantization}\label{Sec:Quantization}

We have presented a simple model that admits a reduced phase space description with residual coordinate freedom. The residual freedom is related to arbitrary local boosts in the observer network associated with our choice of coordinates. Of particular interest is the flat spacetime limit; in this case the momentum equation (\ref{eq:PExact}) can be inverted to obtain the explicit Hamiltonian (\ref{Hflat}), and one could potentially quantize the system exactly by suitably generalizing the quanization of a free relativistic particle.

As a first step, we consider an operator realization of the Hamiltonian (\ref{Hflat}), in the coordinate representation. For convenience, we can split the Hamiltonian into two terms, $H=H_1(x)p+\mathcal{H}(x,p)$, with $H_1(x)=-\sqrt{1-\hat{\lambda}^2}/\hat{\lambda}^2$ and $\mathcal{H}(x,p)=\sqrt{p^2+\hat{M}^2\hat{\lambda}^2}/\hat{\lambda}^2$. The corresponding Schr\"odinger equation $H\Psi(x,t_\lambda)=i\frac{\partial \Psi(x,t_\lambda)}{\partial t_\lambda}$ can be written in time-independent form as
\begin{eqnarray}\label{SchroExact}
-\frac{i}{2}\left( 2H_1\Psi'+H_1'\Psi \right)+\mathcal{H}\Psi=E \Psi,
\end{eqnarray}
with symmetrization of the $H_1(x)p$ term (i.e. $H_1(x)p \Psi\rightarrow -(i/2)(H_1(x)\partial/\partial x + \partial/\partial x H_1(x))\Psi$) chosen to ensure that the Hamiltonian is Hermitian. To define the action of the square root in the $\mathcal{H}\Psi$ term, we will write this term as
\begin{eqnarray}\label{SRTerm}
\mathcal{H}\Psi&=&\left(\frac{\hat{M}}{\hat{\lambda}}\sqrt{1+\left(\frac{p}{\hat{M}\hat{\lambda}}\right)^2}\right)\Psi \nonumber\\
&\rightarrow&\sum_{n=0}^{\infty}c_n p^n \left(\frac{1}{\hat{M}^{2n-1}\hat{\lambda}^{2n+1}}\right)p^n \Psi \nonumber\\
&=& \sum_{n=0}^{\infty}c_n(-1)^n \frac{d^{n}}{dx^{n}}\left(\frac{1}{\hat{M}^{2n-1}\hat{\lambda}^{2n+1}}\frac{d^n \Psi}{\hspace{3pt}dx^n}\right),
\end{eqnarray}
with factor-ordering again chosen for Hermiticity. The expansion coefficients $\{c_n\}$ are trivial to obtain, but will not be needed, as will be clear in what follows.

We will also keep in mind the quadratic momentum limit, given by the arbitrary Hamiltonian (\ref{QuadHamilt}). One can then write the time-independent Schr\"odinger equation as
\begin{eqnarray}\label{SchroGen}
H_0\Psi-\frac{i}{2}\left( 2H_1\Psi'+H_1'\Psi \right)-\left(H_2\Psi'\right)'\approx E \Psi.
\end{eqnarray}
It is clear that the Schr\"odinger equation (\ref{SchroGen}) is merely the truncation of (\ref{SchroExact}) to second order in the $x$-derivatives, with the same symmetrization of both the $H_1p$ term and the $pH_2p$ term.

We can make the scattering region as localized as possible by choosing $M(x<x_\delta)=M_-$ and $M(x>x_\delta)=M_+$, with $M_\pm$ being constants. The tangential pressure barrier is then pointlike, and located at $x_\delta$. We will align a boost with the pressure barrier by taking $\lambda(x<x_\delta)=\lambda_-$ and $\lambda(x>x_\delta)=\lambda_+$, with $\lambda_\pm$ being constants.

For individual modes $e^{ikx}$ in the regions of constant $\hat{M}$ and $\hat{\lambda}$, the time-independent Schr\"odinger equation (\ref{SchroExact}) produces the dispersion relation 
\begin{eqnarray}\label{DispersionExact}
E(k) = -\frac{\sqrt{1-\lambda^2}}{\lambda^2}k+\frac{1}{\lambda^2}\sqrt{k^2+M^2 \lambda^2} .
\end{eqnarray}
On either side of $x_\delta$, there are two independent modes with energy $E$, which correspond to the solutions of (\ref{DispersionExact}), solved for $k$. These modes have wavevectors 
\begin{equation}\label{kExact}
k_{\mp\pm}=E\sqrt{1-\lambda_\pm^2}\mp\sqrt{E^2-M_\pm^2},
\end{equation}
with the indices $\mp$ and $\pm$ chosen independently: $\mp$ to indicate ingoing ($-$) or outgoing ($+$) modes, and $\pm$ to specify the inner ($-$) or outer ($+$) side of the discontinuity. The second-order equation (\ref{SchroGen}) produces the dispersion relation 
\begin{eqnarray}\label{Dispersion}
E(k) \approx H_0+H_1 k+H_2 k^2,
\end{eqnarray}
and a mode with energy $E$ is still associated with multiple wavevectors, this time determined by solving (\ref{Dispersion}) for $k$. The result is
\begin{equation}\label{kWKB}
k_{\mp\pm}\approx -\frac{H_{1\pm}}{2H_{2\pm}}\mp\sqrt{\frac{\left(E-H_{0\pm}\right)}{H_{2\pm}}+\left(\frac{H_{1\pm}}{2H_{2\pm}}\right)^2}.
\end{equation}

The effect of the discontinuity can be determined by integrating (\ref{SchroExact}) across $x_\delta$, which leads to the jump condition
\begin{equation}\label{JumpExact}
\frac{i}{2}\Psi_\delta\left[H_1\right]_\delta=\sum_{n=1}^{\infty}c_n(-1)^n \left[\frac{1}{\hat{M}^{2n-1}\hat{\lambda}^{2n+1}}\frac{d^{2n-1} \Psi}{\hspace{6pt}dx^{2n-1}}\right]_\delta.
\end{equation}
Here $\left[\cdot\right]_\delta$ denotes the jump of a quantity across $x_\delta$, and $\Psi_\delta=\Psi(x_\delta,t_\lambda)$. The truncated form of the jump condition can be obtained by integrating (\ref{SchroGen}), or simply by neglecting all terms but the first on the right-hand side of (\ref{JumpExact}):
\begin{equation}\label{JumpFGen}
\frac{i}{2}\Psi_\delta\left[H_1\right]_\delta\approx-\left[H_2\Psi'\right]_\delta.
\end{equation}

We scatter an initial state localized around $x_0<x_\delta$, which splits into a reflected and transmitted wave when it encounters the pressure barrier. The relevant single-energy scattering modes are characterized by the wavefunction
\begin{displaymath}
   \psi_E(x) = \left\{
     \begin{array}{lr}
       e^{ik_{+-}x}+R_\lambda e^{ik_{--}x} & : x < x_\delta \\
       T_\lambda e^{ik_{++}x} & : x > x_\delta
     \end{array}
   \right.
\end{displaymath}
If we apply wavefunction continuity at $x_\delta$ and the jump condition (\ref{JumpExact}), we can determine the reflection amplitude $R_{\lambda}$ and transmission amplitude $T_{\lambda}$. These amplitudes are given by
\begin{equation}
R_{\lambda}=e^{i(k_{+-}-k_{--})x_\delta}\frac{\chi_{+-}-\chi_{++}-\frac{1}{2}\left[H_1\right]_\delta}{\chi_{++}-\chi_{--}+\frac{1}{2}\left[H_1\right]_\delta}
\end{equation}
and
\begin{equation}
T_{\lambda}=e^{i(k_{+-}-k_{++})x_\delta}\frac{\chi_{+-}-\chi_{--}}{\chi_{++}-\chi_{--}+\frac{1}{2}\left[H_1\right]_\delta},
\end{equation}
with the definition
\begin{equation}
\chi_{\mp\pm}=\frac{\sqrt{k_{\mp\pm}^2+M_\pm^2 \lambda_\pm^2}-M_\pm\lambda_\pm}{\lambda_\pm^2 k_{\mp\pm}}.
\end{equation}
Expressed in terms of energy, $\chi_{\mp\pm}$ can be written as
\begin{equation}
\chi_{\mp\pm}=\frac{\sqrt{1-\lambda_\pm^2}}{\lambda_\pm^2}+\frac{\left(E-\frac{M_\pm}{\lambda_\pm}\right)}{E\sqrt{1-\lambda_\pm^2}\mp\sqrt{E^2-M_\pm^2}}.
\end{equation}

The splitting amplitudes have more transparent forms in the quadratic momentum limit. In this case, the reflection and transmission amplitudes follow from the truncated jump condition (\ref{JumpFGen}), along with continuity at $x_\delta$. If we borrow from the notation for anti-commutators by denoting the sum of a quantity on either side of $x_\delta$ by $\{\cdot\}_\delta$, we can then express the scattering amplitudes compactly as
\begin{equation}
R_{\lambda}\approx -e^{i(k_{+-}-k_{--})x_\delta}\frac{\left[\sqrt{H_1^2+4H_2(E-H_0)}\right]_\delta}{\left\{ \sqrt{H_1^2+4H_2(E-H_0)}\right\}_\delta}
\end{equation}
and
\begin{equation}
T_{\lambda}\approx 2e^{i(k_{+-}-k_{++})x_\delta}\frac{\sqrt{H_{1-}^2+4H_{2-}(E-H_{0-})}}{\left\{ \sqrt{H_1^2+4H_2(E-H_0)}\right\}_\delta}.
\end{equation}

For an initial state, we use a Gaussian wavepacket
\begin{equation}\label{XGaussian}
\Psi(x,0)=\frac{1}{\sqrt{\sigma\sqrt{2\pi}}}e^{i k_0 \left(x-x_0\right)} e^{-\frac{\left(x-x_0\right)^2}{4\sigma^2}},
\end{equation}
with peak location $x_0<x_\delta$, spatial width $\sigma\ll (x_\delta-x_0)$, and initial momentum $k_0$. We can express this initial state in the momentum basis with a Fourier transform
\begin{equation}
\Psi(x,0) =\int_{-\infty}^\infty{\frac{dk}{\sqrt{2\pi}}\Phi(k)e^{ikx}},
\end{equation}
from which we can obtain
\begin{equation}
\Phi(k) = \sqrt{\sigma\sqrt{\frac{2}{\pi}}}e^{-ikx_0}e^{-\sigma^2 (k-k_0)^2}.
\end{equation}

We can alternatively parametrize the single-energy modes by $k=k_{+-}$, and express their wavefunctions as
\begin{eqnarray}
\psi_k(x)&=&\left(e^{ikx}+R_\lambda e^{ik_{--}x}\right)\Theta(x_\delta-x) \nonumber\\
&& +T_\lambda e^{ik_{++}x}\Theta(x-x_\delta),
\end{eqnarray}
with $\Theta$ being the Heaviside step function, and all quantities with energy dependence expressed in terms of $E(k_{+-})=E(k)$. The initial state (\ref{XGaussian}) can be written as an integral over the eigenmodes $\{\psi_k(x)\}$ as
\begin{equation}\label{KModeDecomposition}
\Psi(x,0)=\int\frac{dk}{\sqrt{2\pi}}\Phi(k)\psi_k(x),
\end{equation}
since the contribution from the reflected term is initially localized on the outer side of $x_\delta$ (and so vanishes due to the $\Theta(x_\delta-x)$), and the contribution from the transmitted part is initially localized on the inner side of $x_\delta$ (and so vanishes due to the $\Theta(x-x_\delta)$). 

We can then express the time evolution of the initial state as
\begin{equation}\label{TimeEvolved}
\Psi(x,t_\lambda)=\int\frac{dk}{\sqrt{2\pi}}\Phi(k)\psi_k(x) e^{-iE(k)t_\lambda},
\end{equation}
which involves three terms: one for the incident wavepacket $\Psi_0(x,t_\lambda)$, one for the reflected wavepacket $\Psi_R(x,t_\lambda)$, and one for the transmitted wavepacket $\Psi_T(x,t_\lambda)$. Long after the scattering event, the $\Theta(x_\delta-x)$ factor makes the incident wavepacket vanish; in this case we can drop the $\Theta$ factors in the remaining two terms, since the reflected and transmitted wavepackets are completely free from the pressure barrier.

Before the scattering event, the only contribution to the time-evolved state (\ref{TimeEvolved}) is the first term; if we assume that the initial wavepacket is narrowly peaked in momentum space, we can expand the dispersion relation (\ref{DispersionExact}) to second order about the peak momentum $k_0$. This can be written as
\begin{eqnarray}\label{QuadraticDispersion}
E(k)&\approx& E(k_0) + \frac{d E}{dk_0}\left(k-k_0\right)+\frac{1}{2}\frac{d^2 E}{dk_0^2}\left(k-k_0\right)^2 \nonumber\\
&\equiv& E_0 + v_{g}\left(k-k_0\right)+\beta \left(k-k_0\right)^2,
\end{eqnarray}
with $E_0$ being the peak energy, $v_g$ being the group velocity, and $\beta$ being the dispersion factor of the initial wavepacket. Explicitly, these quantities are
\begin{equation}
E_0 = -\frac{\sqrt{1-\lambda_{-}^2}}{\lambda_{-}^2}k_0+\frac{1}{\lambda_{-}^2}\sqrt{k_0^2+M_{-}^2 \lambda_{-}^2},
\end{equation}
\begin{equation}
v_g = -\frac{\sqrt{1-\lambda_{-}^2}}{\lambda_{-}^2}+\frac{1}{\lambda_{-}^2}\frac{k_0}{\sqrt{k_0^2+M_{-}^2 \lambda_{-}^2}},
\end{equation}
and
\begin{equation}
\beta = \frac{M_{-}^2}{2\left(k_0^2+M_{-}^2 \lambda_{-}^2\right)^{3/2}}.
\end{equation}
The resulting momentum integral is a simple Gaussian, which can be integrated to obtain
\begin{equation}
\Psi_0(x,t_\lambda)=\sqrt{\frac{\sigma}{\sqrt{2\pi}}}\frac{e^{i(k_0(x-x_0)-E_0 t_\lambda)}}{\sqrt{\sigma^2+i\beta t_\lambda}}e^{\frac{-(x-x_0-v_{g}t_\lambda)^2}{4(\sigma^2+i\beta t_\lambda)}}.
\end{equation}

To obtain expressions for the reflection and transmission probabilities $P(R_\lambda)$ and $P(T_\lambda)$, we will take an approach similar to the one presented in \cite{Waves}, whereby the integration variable in the decomposition (\ref{KModeDecomposition}) is converted to the reflected momentum for $P(R_\lambda)$ and the transmitted momentum for $P(T_\lambda)$. First transforming the integration variable to energy, we find
\begin{equation}\label{ReflectedEnergy}
\Psi_R(x,t_\lambda)=\int\frac{dE}{\sqrt{2\pi}}\frac{dk}{dE}\Phi(k(E))Re^{ik_{--}x}e^{-iEt_\lambda}
\end{equation}
and
\begin{equation}\label{TransmittedEnergy}
\Psi_T(x,t_\lambda)=\int\frac{dE}{\sqrt{2\pi}}\frac{dk}{dE}\Phi(k(E))Te^{ik_{++}x}e^{-iEt_\lambda},
\end{equation}
with the conversion factor determined by (\ref{kExact}) to be
\begin{equation}
\frac{dk}{dE}=\sqrt{1-\lambda_{-}^2}+\frac{E}{\sqrt{E^2-M_{-}^2}}.
\end{equation}

We then transform the energy integrals to their respective momentum spaces, using the dispersion relation (\ref{DispersionExact}), as well as the wavevector solutions (\ref{kExact}). This leads to
\begin{equation}\label{ReflectedWavepacketFinal}
\Psi_R(x,t_\lambda)=\int\frac{dk_{--}}{\sqrt{2\pi}}\Phi_R(k_{--})e^{ik_{--}x}e^{-iE(k_{--})t_\lambda}
\end{equation}
and
\begin{equation}\label{TransmittedWavepacketFinal}
\Psi_T(x,t_\lambda)=\int\frac{dk_{++}}{\sqrt{2\pi}}\Phi_T(k_{++})e^{ik_{++}x}e^{-iE(k_{++})t_\lambda},
\end{equation}
with
\begin{equation}
\Phi_R(k_{--})=\frac{dk}{dE}\frac{dE}{\hspace{8pt}dk_{--}}\Phi(k(E(k_{--})))R_\lambda
\end{equation}
and
\begin{equation}
\Phi_T(k_{++})=\frac{dk}{dE}\frac{dE}{\hspace{8pt}dk_{++}}\Phi(k(E(k_{++})))T_\lambda.
\end{equation}
The transformation factors that convert between energy and the momentum spaces are found from (\ref{DispersionExact}) to be
\begin{equation}
\frac{dE}{\hspace{8pt}dk_{\pm\pm}}=-\frac{\sqrt{1-\lambda_{\pm}^2}}{\lambda_{\pm}^2}+\frac{1}{\lambda_{\pm}^2}\frac{k_{\pm\pm}}{\sqrt{k_{\pm\pm}^2+M_{\pm}^2 \lambda_{\pm}^2}}.
\end{equation}

At this point Parseval's theorem can be exploited to determine the reflection and transmission probabilities from (\ref{ReflectedWavepacketFinal}) and (\ref{TransmittedWavepacketFinal}), resulting in
\begin{equation}\label{RefProbExact}
P(R_\lambda)=\int d k_{--} |\Phi_R(k_{--})|^2
\end{equation}
and 
\begin{equation}\label{TransProbExact}
P(T_\lambda)=\int d k_{++} |\Phi_T(k_{++})|^2,
\end{equation}
respectively. The advantage of this approach is clear: it allows the scattering probabilities to be expressed as time-independent integrals over the respective momentum spaces. The cost of this simplification is that the integrands are highly nontrivial. 

We can obtain more enlightening expressions for the reflection and transmission probabilities by transforming the integration variable back to $k$, as pointed out in \cite{Waves}. This leads to
\begin{equation}\label{ReflectionExact}
P(R_\lambda)=\int dk|\Phi(k)|^2 \left( \frac{dk}{dE}\frac{dE}{\hspace{8pt}dk_{--}} \right)|R_\lambda(k)|^2
\end{equation}
and
\begin{equation}\label{TransmissionExact}
P(T_\lambda)=\int dk|\Phi(k)|^2 \left( \frac{dk}{dE}\frac{dE}{\hspace{8pt}dk_{++}} \right) |T_\lambda(k)|^2,
\end{equation}
with $k_{--}$ and $k_{++}$ understood to be functions of $k$, as determined by (\ref{kExact}). The expressions (\ref{ReflectionExact}) and (\ref{TransmissionExact}) demonstrate that the reflection/transmission probabilities for a wavepacket can be represented as an integral over incoming momentum modes, with each contribution given by the reflection/transmission probability for a single mode multiplied by the probability density for the mode to be part of the initial state.

Converting the reflected and transmitted wavepacket integrals to their respective momentum spaces is much more transparent in the quadratic limit. For this purpose, it is useful to write the wavevector solutions (\ref{kWKB}) as $k_{\mp\pm}\approx k_{0\pm}\mp k_\pm$. The reflected wavevector is then $k_{--}\approx k_{0-}-k_-$, and since the incident wavevector is $k_{+-}\approx k_{0-}+k_-$, we can convert the reflected wavepacket integral with $k\approx 2k_{0-}-k_{--}$. This leads to the reflected wavepacket expression
\begin{equation}
\Psi_R(x,t)\approx \int\frac{d k_{--}}{\sqrt{2\pi}}R_\lambda \Phi(2k_{0-}-k_{--})e^{i k_{--}x}e^{-iE t_\lambda},
\end{equation}
where all quantities with energy dependence are understood to be expressed in terms of $E(k_{--})$. We are also now assuming we have waited long enough after the scattering event to omit the $\Theta$ factors.

We can now make use of Parseval's theorem again to write the reflection probability as
\begin{equation}\label{RefProb}
P(R_\lambda)\approx \int d k_{--} |\Phi_R(k_{--})|^2,
\end{equation}
for
\begin{equation}
\Phi_R(k_{--})=R_\lambda\Phi(2k_{0-}-k_{--}).
\end{equation}
No time-dependence is present in the integral (\ref{RefProb}), though even in this limit the integrand is nontrivial. We note that due to the $(2k_{0-}-k_{--})$ that appears in the Gaussian function $\Phi$, we will choose our parameters such that only negative $k_{--}$ contribute to the integral (\ref{RefProb}), as must be the case for a reflected wavepacket.

The transmission probability requires more care to calculate. Since $k_{++}\approx k_{0+}+k_+$, and the original expression (\ref{TimeEvolved}) involves integration over $k=k_{+-}\approx k_{0-}+k_-$, transforming the incident momentum integral to the transmitted momentum integral gives us
\begin{eqnarray}
\Psi_T(x,t_\lambda)&\approx &\int \frac{d k_{++}}{\sqrt{2\pi}}T_\lambda \Phi(k_{0-}+k_-(E)) \\
&&\cdot \frac{\sqrt{H_{1+}^2+4H_{2+}(E-H_{0+})}}{\sqrt{H_{1-}^2+4H_{2-}(E-H_{0-})}}e^{i k_{++} x}e^{-iEt_\lambda}, \nonumber
\end{eqnarray} 
with all energy dependence expressed in terms of $E(k_{++})$. The transmission probability is then
\begin{equation}
P(T_\lambda)\approx \int d k_{++} |\Phi_T(k_{++})|^2,
\end{equation}
for
\begin{eqnarray}
\Phi_T(k_{++})&=&T_\lambda \Phi(k_{0-}+k_-(E))\\
&&\cdot\frac{\sqrt{H_{1+}^2+4H_{2+}(E-H_{0+})}}{\sqrt{H_{1-}^2+4H_{2-}(E-H_{0-})}}.\nonumber
\end{eqnarray}

Transforming back to integration over $k$, we find
\begin{equation}\label{Reflection}
P(R_\lambda)\approx \int dk|\Phi(k)|^2 |R_\lambda(k)|^2
\end{equation}
and
\begin{equation}\label{Transmission}
P(T_\lambda)\approx \int dk|\Phi(k)|^2|T_\lambda(k)|^2 \frac{\left(2H_{2+}k_{++}+H_{1+}\right)}{\left(2H_{2-}k+H_{1-}\right)},
\end{equation}
with $k_{++}$ understood to be a function of $k$, as determined by (\ref{kWKB}).

The $k$-parametrization also enables us to re-express $|R_\lambda|^2$ and $|T_\lambda|^2$ as
\begin{equation}
|R_\lambda|^2\approx \frac{\left(2H_{2+}k_{++}-2H_{2-}k+\left[H_1\right]_\delta\right)^2}{\left(2H_{2+}k_{++}+2H_{2-}k+\{H_1\}_\delta\right)^2}
\end{equation} 
and
\begin{equation}
|T_\lambda|^2\approx \frac{4\left(2H_{2-}k+H_{1-}\right)^2}{\left(2H_{2+}k_{++}+2H_{2-}k+\{H_1\}_\delta\right)^2}.
\end{equation} 

\begin{figure}
\centering
  \includegraphics[width=0.5\textwidth]{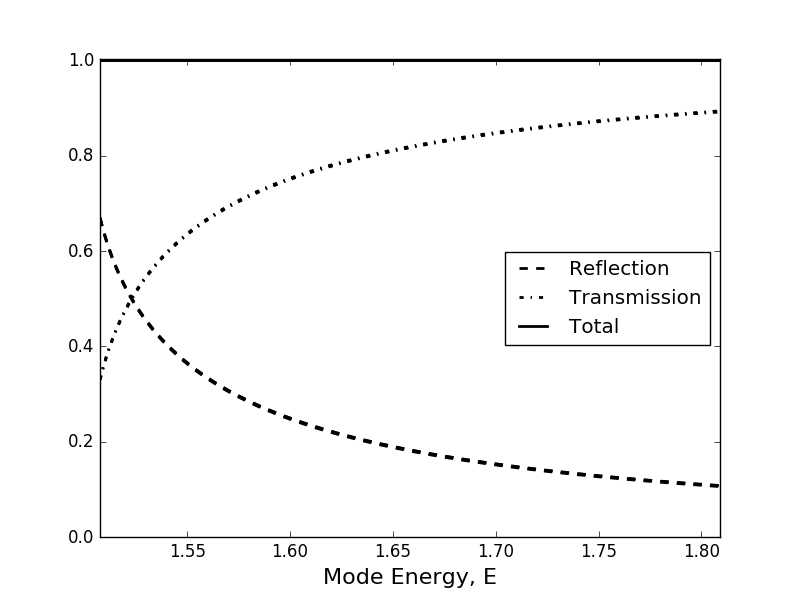}
  \caption{Comparison of single-mode reflection and transmission probabilities, (\ref{NarrowR}) and (\ref{NarrowT}), plotted against the incident wave energy $E$. The parameters are $M_-=1$, $M_+=1.5$, $v_-=0$, and $v_+\in [0,0.1]$.}
	\label{fig:QuadSplit}
\end{figure}

For sufficiently narrow momentum-space distributions, we can evaluate the non-exponential factors of the integrands in the reflection and transmission probabilities at the peak momentum $k_0$. In this case we can recover the single-mode reflection probability
\begin{eqnarray}\label{NarrowR}
P(R_\lambda)&\approx& |R_\lambda(k_0)|^2\sigma \sqrt{\frac{2}{\pi}}\int dk e^{-2\sigma^2(k-k_0)^2} \nonumber\\
&=& |R_\lambda(k_0)|^2,
\end{eqnarray}
as well as the single-mode transmission probability as
\begin{eqnarray}\label{NarrowT}
P(T_\lambda)&\approx& |T_\lambda(k_0)|^2\frac{\left(2H_{2+}\bar{k}_{++}+H_{1+}\right)}{\left(2H_{2-}k_0+H_{1-}\right)} \nonumber\\
&& \cdot\sigma\sqrt{\frac{2}{\pi}}\int dk e^{-2\sigma^2(k-k_0)^2}\nonumber\\
&=& |T_\lambda(k_0)|^2\frac{\left(2H_{2+}\bar{k}_{++}+H_{1+}\right)}{\left(2H_{2-}k_0+H_{1-}\right)},
\end{eqnarray}
with $\bar{k}_{++}$ being the peak transmitted momentum. The single-mode splitting probabilities are plotted in Figure \ref{fig:QuadSplit}. With a vanishing inner observer network velocity ($v_-=\sqrt{1-\lambda_-^2}=0$), several different outer observer velocities up to $10\%$ of the speed of light were included, with nearly perfect overlap. We conclude, then, that scattering probabilities in flat spacetime are unaffected by local boosts to the observer network, for nonrelativistic speeds.  

\begin{figure*}
\centering
  \includegraphics[width=0.99\textwidth]{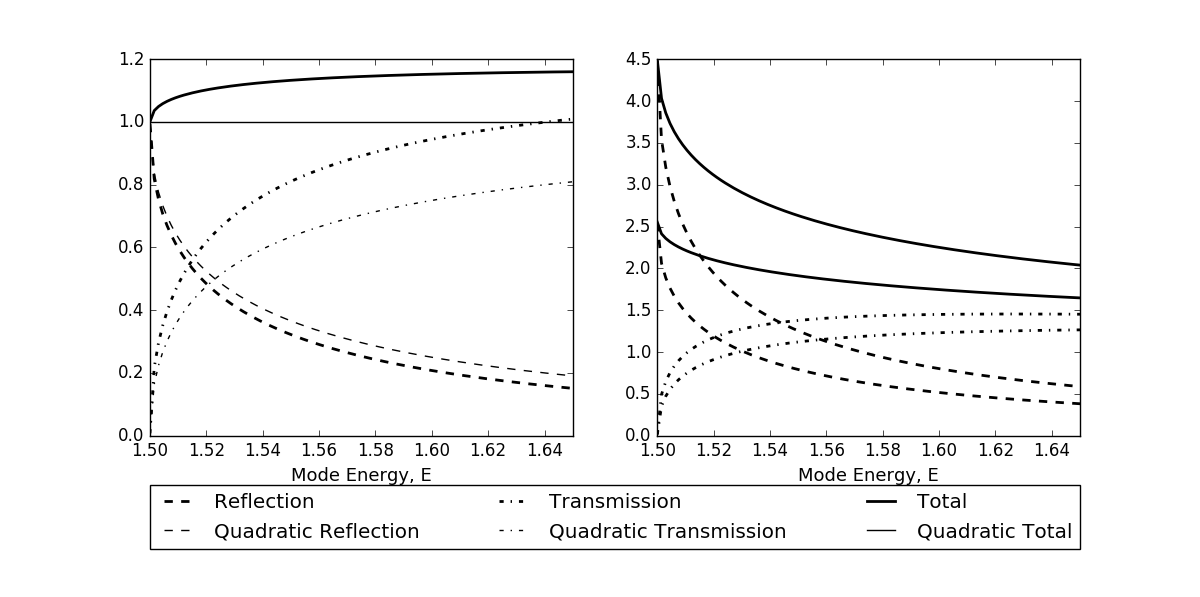}
  \caption{Comparison of single-mode limits of the exact flat spacetime reflection and transmission probabilities, (\ref{ReflectionExact}) and (\ref{TransmissionExact}), plotted against the incident wave energy $E$, with $M_-=1$, $M_+=1.5$, and $v_-=0$. In the plot shown on the left (for which $v_+=0$), the exact probabilities are contrasted with the quadratic limit splitting probabilities. The plot on the right shows the exact flat spacetime splitting probabilities for $v_+=0.2$ (lower curves) and $v_+=0.3$ (upper curves). The failure to recover consistent single-mode splitting as relativistic energies are approached is evidenced by the reflection and transmission probabilities summing to more than unity.}
	\label{fig:Split}
\end{figure*}

As a simple test of these results, consider nonrelativistic scattering off of a step potential, with $H_{0-}=m$, $H_{1\pm}=0$, $H_{0+}=m+V_0$, and $H_{2\pm}=1/2m$. Then $k_{+-}=k=\sqrt{2m(E-m)}$, $k_{++}=\sqrt{2m(E-m-V_0)}$, and we can express the reflection and transmission probabilities in the narrow momentum-space distribution limit as
\begin{equation}
P(R)=\frac{\left(k_0-\bar{k}_{++}\right)^2}{\left(k_0+\bar{k}_{++}\right)^2}
\end{equation}
and
\begin{equation}
P(T)=\frac{4k_0\bar{k}_{++}}{\left(k_0+\bar{k}_{++}\right)^2},
\end{equation}
respectively. These expressions agree with the standard textbook result, as quoted, for instance, in \cite{Waves}.

Another natural consistency check is to test whether single-mode probabilities can be recovered from the exact flat spacetime wavepacket splitting probabilities (\ref{ReflectionExact}) and (\ref{TransmissionExact}) in the limit that the wavepackets are very narrowly-peaked in momentum space. Surprisingly, this turns out to be problematic. Figure \ref{fig:Split} illustrates the clear departure of the relativistic amplitudes from the probabilities derived in the limit of quadratic momentum, for all but the lowest allowable speeds (since we wish to describe scattering, the lowest incident energy is $\text{max}\{M_-,M_+ \}$). The effect is much less pronounced for smaller mass gaps - which correspond to small pressure barriers - and the problem disappears as the barrier height and outer observer network boosting tend to zero. We discuss the origins of this problem in the following section, and offer suggestions for a solution.

\section{Discussion}\label{Sec:Discussion}

As shown in Figure \ref{fig:Split}, the quantization scheme considered here has a problem with probability nonconservation. The problem is not present in the nonrelativistic limit (i.e. the quadratic momentum limit), but with the exception of the trivial case where both pressure barrier and network boosting vanish, it is present everywhere else in the parameter space. This issue arises due to a conflict between the probability density definition $\rho(x,t)=|\Psi(x,t)|^2$ and our Hamiltonian operator realization;
although formally Hermitian, the factor ordering (\ref{SRTerm}) does not yield a self-adjoint Hamiltonian.

We can see exactly where the problem lies by starting with the the condition of probability conservation, $(d/dt)\int dx\, |\Psi(x,t)|^2=0$. Taking the time derivative inside the integral and applying the Schr\"odinger equation yields 
\begin{equation}\label{SelfAdj}
\int dx\, \Psi^* (H\Psi)=\int dx\, \Psi(H\Psi)^*,
\end{equation}
which is the condition that $H$ is a self-adjoint operator. With the factor ordering (\ref{SRTerm}), the left hand side of (\ref{SelfAdj}) can be integrated by parts term by term to verify consistency with the right hand side. The $n=0$ verification is trivial; at $n=1$, we find
\begin{align}
&\int dx\, \left(\Psi^* p\left(\frac{1}{\hat{M}\hat{\lambda}^3}\right) p\Psi - \Psi \left[p\left(\frac{1}{\hat{M}\hat{\lambda}^3}\right) p\Psi\right]^*\right)\nonumber\\
&= \left[\frac{1}{\hat{M}\hat{\lambda}^3}\left(\Psi\partial_x\Psi^*-\Psi^*\partial_x\Psi\right)\right]_\delta,\label{n1}
\end{align}
assuming $\Psi$ is localized sufficiently close to the scattering region near $x=x_\delta$. Using continuity of the wavefunction at $x_\delta$, the boundary contribution in (\ref{n1}) can be written as
\begin{align}
\Psi_\delta \left[\frac{1}{\hat{M}\hat{\lambda}^3}\partial_x \Psi^* \right]_\delta-\Psi_\delta^*\left[\frac{1}{\hat{M}\hat{\lambda}^3}\partial_x \Psi \right]_\delta.
\end{align}
This contribution is consistent with the $n=1$ term in (\ref{JumpExact}), which is also present in the truncated jump condition (\ref{JumpFGen}).

The $n=2$ term does not maintain this consistency. In this case we have 
\begin{align}
&\int dx\, \left(\Psi^*  p^2\left(\frac{1}{\hat{M}^3\hat{\lambda}^5}\right)p^2\Psi - \Psi \left[p^2\left(\frac{1}{\hat{M}^3\hat{\lambda}^5}\right)p^2\Psi\right]^*\right)\nonumber\\
&=\left[\frac{1}{\hat{M}^3\hat{\lambda}^5}\left(\Psi^*\partial_x^3\Psi-\Psi\partial_x^3\Psi^*\right)\right]_\delta\nonumber\\
&\hspace{12pt}+\left[\frac{1}{\hat{M}^3\hat{\lambda}^5}\left(\partial_x^2\Psi^*\partial_x\Psi-\partial_x^2\Psi\partial_x\Psi^*\right)\right]_\delta\label{n2}
\end{align}
As with the $n=1$ case, the first boundary contribution in (\ref{n2}) is consistent with the $n=2$ term in (\ref{JumpExact}), assuming continuity of the wavefunction at $x_\delta$. The second boundary contribution in (\ref{n2}), involving the quantity $(\partial_x^2\Psi^*\partial_x\Psi-\partial_x^2\Psi\partial_x\Psi^*)$, has no counterpart in the derived jump condition (\ref{JumpExact}); herein lies the source of the probability nonconservation encountered above.  

The inconsistency between our jump condition (\ref{JumpExact}) and the condition of self-adjointness (\ref{SelfAdj}) for our Hamiltonian, given the factor ordering (\ref{SRTerm}), implies that one cannot both enforce wavefunction continuity and derive a jump condition by integrating the Schr\"odinger equation across $x_\delta$, if one also wishes to choose the probability density $\rho(x,t)=|\Psi(x,t)|^2$. A natural option to resolve the issue would then be to choose a jump condition based on self-adjointness (\ref{SelfAdj}) rather than by integrating the Schr\"odinger equation across $x_\delta$. This option is compatible with the method of self-adjoint extensions \cite{SelfAdjExt}\cite{OpDomains}. With this method, an inner-product is chosen together with wavefunction boundary conditions such that the operator realization of the Hamiltonian is manifestly self-adjoint - this ensures the associated scattering process conserves probability. Applying the method of self-adjoint extensions to the model presented here is a technical challenge even in the flat spacetime limit; further research is required to determine whether this method is suitable for quantizing the reduced shell system.

Another potential option to define a consistent quantization for our system is ``second quantization''. Similar to obtaining the Klein-Gordon equation by ``squaring'' the operator form of $H=\sqrt{p^2+m^2}$ and applying it to a field, one might hope to obtain a sensible theory by squaring the operator form of the analogous square root expression in (\ref{Hflat}). Such a procedure may seem ad-hoc, but would have the benefit of allowing pair-production effects to be taken into account, which may be crucial to understanding how coordinate choices affect the scattering behaviour.

It is remarkable that due to the equivalence of the ADM mass and the reduced Hamiltonian of our system, different members of the Painlev\'e-Gullstrand family (as well as hybrids of them) measure time differently, but all share the same Hamiltonian, even in the flat spacetime limit. For global boosts (in one dimension) between an inertial frame $F$ and an inertial frame $\tilde{F}$ moving at a velocity $v$ with respect to $F$, the usual special relativistic transformation of energy and momentum is $\left(\tilde{E},\tilde{p}\right)=\left(\gamma (E-vp),\gamma (p-vE)\right)$; in our case, we can see from the form of (\ref{eq:PExact}) that the transformation between the Painlev\'e-Gullstrand frame ($\lambda=1$) and an arbitrary frame in the Painlev\'e-Gullstrand family ($0<\lambda<1$) is given by $\left(E_\lambda,p_\lambda\right)=\left(E,p+vE\right)$ (keeping in mind that $v>0$ indicates inward motion of the observer network associated with the frame). However, the difference between these transformations does not indicate an inconsistency, for two reasons. The first is that our frames are not related by global boosts. The second reason is that though the spatial slices in the $\lambda\neq 1$ frame differ from those in the $\lambda=1$ frame (the former have curvature and the latter do not), there is no motion between a constant-$r$ point of one frame and a constant-$r$ point of the other. It is only the time coordinates that reflect the infalling nature of the associated observer networks.

It remains an open question what the exact connection is between coordinate choices and observers in quantum gravity. We do know that different choices of the time variable can lead to unitarily inequivalent matter field quantizations, which affects how observers interpret the particle content of a field. It is not known in general what coordinate systems that lead to unitarily equivalent quantizations have in common, but one might naturally think it is related to shared inertial features of the associated observer networks.

If the model presented here can be consistently quantized, we will have a way to directly probe the effect a locally non-inertial observer network has on the corresponding quantization, and potentially resolve some outstanding related issues. Is the local velocity $\hat{v}$ associated with the observer network purely a feature of how we label spacetime events, or can it be interpreted as an actual degree of freedom? In spherical symmetry, there are no propagating gravitational degrees of freedom in the usual sense, but a given coordinate system can still define inertial and non-inertial regions. For non-inertial regions, is additional structure required to obtain a sensible interpretation, such as the introduction of dynamical variables to represent the sources of non-inertial observer motion? If so, can observer networks themselves exhibit quantum features?

\section{Acknowledgements}

The authors would like to thank the Natural Sciences and Engineering Research Council of Canada (NSERC) and the Templeton Foundation (Grant No. JTF $36838$) for financial support. CG acknowledges additional support from Silke Weinfurtner at the University of Nottingham. The authors are also grateful to the Aspelmeyer and Brukner groups at the University of Vienna, as well as Friedemann Queisser and Jorma Louko, for stimulating discussions.


\begin{thebibliography}{99}

\bibitem{GoodingUnruh14}
  C.~Gooding and W.~G.~Unruh,
	``Self-gravitating interferometry and intrinsic decoherence,''
	Phys.\ Rev.\  D {\bf 90}, 044071 (2014).

\bibitem{PainleveFamily}
  K.~Martel and E.~Poisson,
  ``Regular coordinate systems for Schwarzschild and other spherical spacetimes,'' 
  Am.\ J.\ Phys.\ {\bf 69}, 476 (2001).
  [arXiv:gr-qc/0001069v4]

\bibitem{Menotti}
	F.~Fiamberti and P.~Menotti,
  ``Reduced Hamiltonian for intersecting shells,''
	Nucl.\ Phys.\ B {\bf 794}, 512 (2008).
  [arXiv:hep-th/0708.2868v1]

\bibitem{ADM}
	R.~L.~Arnowitt, S.~Deser, and C.~W.~Misner,
  ``Canonical Variables for General Relativity,''
  Phys.\ Rev.\ {\bf 117}, 1595 (1960).
  [arXiv:gr-qc/0405109]	

\bibitem{Penrose65}
  	R.~Penrose,
	``Gravitational Collapse and Space-Time Singularities,''
	Phys.\ Rev.\ Lett.\ {\bf 14}, 57 (1965).	
	
\bibitem{Louko}
  J.~L.~Friedman, J.~Louko, and S.~N.~Winters-Hilt,
	``Reduced phase space formalism for spherically symmetric geometry with a massive dust shell,''
	Phys.\ Rev.\ D {\bf 56}, 7674 (1997).
	
\bibitem{Waves}
  T.~Norsen, J.~Lande, and S.~B.~McKagan,
	``How and why to think about scattering in terms of wave packets instead of plane waves,''
	(unpublished), (2009).
	[arXiv:0808.3566v2]
	
\bibitem{SelfAdjExt}
  G.~Bonneau, J.~Faraut, and G.~Valent,
	``Self-adjoint extensions of operators and the teaching of quantum mechanics,''
	Am.\ J.\ Phys.\ {\bf 69}, 322 (2001).
	
\bibitem{OpDomains}
  V.~S.~Araujo, F.~A.~B.~Coutinho, and J.~Fernando Perez,
	``Operator domains and self-adjoint operators,''
	Am.\ J.\ Phys.\ {\bf 72}, 203 (2004).	

\end{thebibliography}
\end{document}